\documentclass{article}

\usepackage{arxiv}

\usepackage[utf8]{inputenc} 
\usepackage[T1]{fontenc}    
\usepackage{booktabs}       
\usepackage{amsfonts}       
\usepackage{graphicx}
 \usepackage{amsmath} 

\title{Exploration of the potential energy surface for the conformational interconversion of the amyloid $\beta$ peptide at the fibril end}

\author{
 Yasuhiro Oishi \\
  Graduate School of Science\\
  University of Hyogo\\
  3-2-1 Koto, Kamiri-cho, Ako-gun, 678-1297, Japan \\
  \texttt{rk23m002@guh.u-hyogo.ac.jp} \\
   \And
 Motoharu Kitatani \\
   Graduate School of Science\\
  University of Hyogo\\
  3-2-1 Koto, Kamiri-cho, Ako-gun, 678-1297, Japan \\
  \texttt{kitatani@sci.u-hyogo.ac.jp} \\
  \And
 Kichitaro Nakajima \\
  Graduate School of Engineering\\
  Osaka University\\
  2-1 Yamadaoka, Suita, Osaka, 565-0871, Japan \\
  \texttt{k.nakajima@prec.eng.osaka-u.ac.jp} \\
    \And
 Hirotsugu Ogi \\
  Graduate School of Engineering \\
  Osaka University\\
  2-1 Yamadaoka, Suita, Osaka, 565-0871, Japan \\
  \texttt{ogi@prec.eng.osaka-u.ac.jp} \\
    \And
 Koichi Kusakabe \\
  Graduate School of Science\\
  University of Hyogo\\
  3-2-1 Koto, Kamiri-cho, Ako-gun, 678-1297, Japan \\
  \texttt{kusakabe@sci.u-hyogo.ac.jp} \\
}

\begin{document}
\maketitle
\begin{abstract}
The formation of amyloid fibrils comprising amyloid $\beta$ (A$\beta$) peptides is associated with the pathology of Alzheimer's disease.
In this study, we theoretically investigated the A$\beta$ structure at the fibril end using the density functional theory calculation. 
Several twisted conformations were identified as local minima in which a part of the peptide chain bends upward while the rest remains bound to the lower A$\beta$ monomer.
Fibril-to-twisted conformational transition exhibited endothermic behavior, with endothermic energy increasing as more backbone hydrogen bonds were broken.
In addition, the loss of van der Waals interaction from the hydrophobic sidechain contributed to endothermicity.
The nudged elastic band method was applied to analyze the potential energy surface connecting the fibril and twisted conformations.
Comparison of the activation barriers between different twisted conformations revealed that certain twisted conformations returned relatively easily to the fibril conformation, whereas others encountered a higher activation barrier and reverted less readily.
Detailed structural analysis revealed that the twisted conformation's propensity to return originates from the local steric hindrance imposed by the sidechain near the torsional axis. 
\end{abstract}

\section{Introduction}
Protein aggregation often results in the formation of amyloid fibrils, which exhibit a characteristic needle-like morphology [1] and play a crucial role in the amyloidosis pathology [2]. In particular, Alzheimer's disease is associated with amyloid fibrils comprising amyloid $\beta$ (A$\beta$) peptides [3,4]. The A$\beta$ peptide is generated through the cleavage of the amyloid precursor protein by $\beta-$ and $\gamma-$secretases [5,6], and the generated peptide exhibits a tendency to self-assemble, finally forming fibrillar structures.

 Amyloid fibril structures have been extensively characterized by experimental techniques, such as X-ray diffraction [7,8], solid-state nuclear magnetic resonance [9-14], and Cryogenic Electron Microscopy [15-19]. In the bulk region of the fibril, the A$\beta$ adopts a cross-$\beta$ structure and forms parallel in-register $\beta$-sheets with adjacent monomers, constituting a one-dimensional fibrillar morphology. In contrast, at the terminal region of the fibril, the A$\beta$ conformation may differ from that in the bulk region [20]. The elongation process of the fibril proceeds via a consecutive addition of the monomer to the fibril end [21]. Therefore, gaining deeper insight into the fibril formation mechanism requires understanding the detailed conformational characteristics of A$\beta$ at the fibril end.
 
Theoretically, the energetics of amyloid-like peptide molecules have been extensively investigated using the density functional theory (DFT) to accurately describe the hydrogen bond (HB) interactions [22-24]. These studies have primarily focused on elucidating how HBs contribute to amyloid fibril stability and formation and have examined the association and binding energy of peptides in crystalline forms. However, single-molecule conformational changes in the amyloid fibril remain theoretically unexplored.

A possible mode of such conformational changes in protein molecules is the torsional motion around the backbone dihedral angles\rule[0.5ex]{0.2cm}{0.4pt}that is\rule[0.5ex]{0.2cm}{0.4pt}the Ramachandran dihedral angle of $\psi$ and $\phi$ defined for the C-C and N-C single bonds in the peptide unit. The potential energy surface (PES) associated with these torsions has been explored in other peptide systems [25-28].

In this study, we theoretically investigate the torsional motion of the A$\beta$ at the fibril end. The energetics of this motion is explored using the nudged elastic band (NEB) method based on DFT. The stability and structural features resulting from torsion are examined in detail. We reveal the role of local steric hindrance in creating an activation barrier associated with torsional motion. Based on these findings, we discuss how the local structural feature can affect the entire monomeric structure of the A$\beta$ at the terminal region of the amyloid fibril.

\section{Computational methods and materials}
\subsection{Structural model for the fibril end}
We investigated the structural changes at the terminal region of the amyloid fibril. As a structural model representing the fibril end, we employed a dimer model in which two monomers were aligned in a parallel $\beta$-sheet conformation. As a monomer, the structure of the crystallized peptide encoding residues 20-34 of the A$\beta$ peptide, determined by the electron diffraction experiment [29], was employed in the present study. The atomic coordinates of the crystal structure were extracted from the Protein Data Bank (PDB ID: 6OIZ).

We focused on the torsional motions around single bonds within the monomer located at the fibril end. 
The direction of torsion and the possible torsional axis are limited due to intermolecular and intramolecular steric hindrances. 
Some torsions result in a structure in which part of the upper monomer's peptide chain separates from the lower monomer, while the remaining segment remains stacked through a parallel $\beta-$sheet. We define such a structure as a twisted conformation. 
In contrast, we define a structure in which the upper monomer is fully stacked on the lower monomer via a parallel $\beta-$sheet as a fibril conformation.

\subsection{Potential energy surface calculation}
The DFT calculation was implemented using the PWscf code of Quantum ESPRESSO [30-32]. For the exchange-correlation functional, the PBEsol functional [33] together with the DFT-D3 correlation [34] was used to describe the van der Waals interaction. Ultrasoft pseudopotentials [35] with an energy cutoff of 25(250) Ry for wavefunction (charge density) expansion were employed. The calculation was performed only at the gamma-point for an isolated dimer model in a cell with a size of $45\times45\times40$ \AA.
Structural optimization was performed for the fibril and twisted conformation until the Hellman-Feynman force acting on each atom was less than $10^{-5}$ Ry/Bohr. The NEB method [36] was employed to search for the pathway connecting the fibril and twisted conformation. We used Xcrysden [37] to measure the interatomic distance and display the atomic structure.

\section{Results}
This section presents the structural and energetic analysis of the A$\beta$ dimer.
The structure of the fibril conformation is first described, after which the twisted conformations and their relative energies to the fibril conformation are presented.
The energy trend among the different twisted conformations is then explained based on their structural differences.
In addition, the PESs connecting the fibril and twisted conformations using the NEB calculation are presented.
Activation barriers associated with each torsion are compared, and the origins of the barrier heights are analyzed.

\subsection{Structure of the fibril conformation}

Figure 1 shows the optimized structure of the fibril conformation. The two monomers are tightly connected through 13 HBs, of which 12 involve mainchain atoms and one involves the sidechains of N27 residue. As an intermolecular HB, we adopted the criteria for HB employed in the literature [38]; a HB was considered present if the distance between the donor atom (D) and acceptor atom (A) was less than or equal to 3.6 \text{\AA}, and the D–H···A angle was greater than or equal to $120^\circ$.

 The A$\beta$ peptide exhibits a nearly periodic meandering polypeptide structure; bends constituting the meandering are formed in every segment containing three consecutive bonds (C-C, CO-NH, and N-C bonds). A segment containing the boundary between the i-th and i+1-th amino acid residues is called an i-i+1 segment.

\subsection{Structure and stability of the twisted conformation}

Figure 2 shows the optimized structures of the twisted conformation. Each twisted conformation is labeled according to the axis around which the structure is twisted using the residue and Ramachandran angle. For example, “$\psi$ torsion of E22” refers to torsion around the C-C bond at the E22 residue.

Figure 3 summarizes the energies of the twisted conformations relative to the fibril conformation, showing that all twisted conformations are less stable than the fibril conformation. The energy of the twisted conformations increased with the length of the peptide chain lifted under torsion, except for the $\phi$ torsion of D23.

For each twisted conformation, all intermolecular mainchain HBs located from the torsional axis to the N-terminus in the fibril conformation were broken. The number of broken HBs is denoted as $\Delta \rm N_{\rm HB}$ and is shown at the top of each bar graph in Figure 3. A roughly proportional relationship between endothermic energy and $\Delta \rm N_{\rm HB}$ was observed, except for the $\phi$ torsion of D23.

Here, we focus on the energy difference between $\psi$ and $\phi$ torsions within the same segment ($\psi$ torsion of E22 vs $\phi$ torsion of D23, $\psi$ torsion of D23 vs $\phi$ torsion of V24, and $\psi$ torsion of G25 vs $\phi$ torsion of S26). For each segment, the endothermic energy of the twisted conformation caused by $\phi$ torsion exceeded that caused by $\psi$ torsion. The common reason for this is that the $\Delta \rm N_{\rm HB}$ is one more in $\phi$ torsion than in $\psi$ torsion for each segment. The energy difference between $\phi$ and $\psi$ torsion was 0.64, 0.32, and 0.26 eV for the 22-23, 23-24, and 25-26 segments, respectively.

The largest energy difference in the 22-23 segment (0.64 eV) arises from repulsion between the mainchain carbonyl oxygen atoms of the upper and lower monomers in the 22-23 segment. Unlike the $\psi$ torsion, for the $\phi$ torsion, the oxygen atom in the 22-23 segment of the upper monomer approaches the oxygen atom in the same segment of the lower layer. In the $\phi$ torsion, the distance between the oxygen atoms is 4.7783 \text{\AA} in the fibril conformation, whereas in the twisted conformation, it is reduced to 2.9932 \text{\AA}. This proximity seems to destabilize the twisted conformation caused by the $\phi$ torsion of D23.

Another type of interaction, such as the van der Waals interaction, also affects endothermic energy.
This notably reflects two energy trends: the endothermic energy of the $\psi$ torsion of E22 and the energy difference between the $\phi$ torsion of V24 and the $\psi$ torsion of G25.
First, we discuss the $\psi$ torsion of E22.
The endothermic energy is 1.12 eV despite only one HB being broken in this torsion.
The relationship between endothermic energy and $\Delta \rm N_{\rm HB}$ indicates that the torsion-induced energy changes cannot be explained solely by $\Delta \rm N_{\rm HB}$. 
All calculated torsions involve the breakage of $\pi$-stacking of the F20 sidechain, stabilizing the fibril conformation via the van der Waals interaction.
This breakage partly explains why the twisted conformation arising from the $\psi$ torsion of E22 has such a high energy, even though only one HB is broken.
Second, the energy difference between the $\phi$ torsion of V24 and the $\psi$ torsion of G25 reaches 0.92 eV, even though the corresponding difference in the $\Delta \rm N_{\rm HB}$ is one.
A possible explanation to this is the van der Waals interaction arising from the sidechain of V24. 
In the fibril conformation, the V24 sidechain of the upper monomer faces the sidechains of hydrophobic residues of the upper monomer, such as I31, as well as the V24 sidechain of the lower monomer. These contacts may enhance the stability of the fibril conformation through van der Waals interactions. In the $\psi$ torsion of G25, the V24 sidechain in the upper monomer is separated from these sidechains, unlike in the $\phi$ torsion of V24. Thus, the loss of the van der Waals interaction that stabilizes the fibril conformation is greater in the $\psi$ torsion of G25 than in the $\phi$ torsion of V24, which seems to increase the energy difference.

Overall, the intermolecular HB and the van der Waals interaction from the hydrophobic sidechain stabilized the fibril conformation.
The results thus far indicate that when a torsional motion disrupts a greater number of such interactions, the resulting twisted conformation becomes less stable, thereby increasing endothermic energy. Additionally, in torsions where the carbonyl oxygen atoms in the upper and lower monomers approach, the twisted conformation becomes further destabilized. 
These factors make the transition from the fibril to the twisted conformation less probable.
Although our study focuses on torsions that lead to the lifting of the peptide chain from the N-terminus, similar results are expected for torsions that lift the peptide chain from the C-terminus.

\subsection{Pathway and activation barrier for torsional motion}

Here, we discuss the reaction pathway connecting the fibril and twisted conformations obtained using the NEB calculation.
Figure 4 shows the PES connecting the fibril and twisted conformations, with the energy of the fibril conformation taken as the reference.
All six torsions exhibited endothermic behavior.
First, we focus on the difference in the curve of PES between $\psi$ and $\phi$ torsions within the same segment. For the 22-23 segment, the energy increase was larger for the $\phi$ torsion of D23 than for the $\psi$ torsion of E22, arising from the proximity between the oxygen atoms, as discussed previously. For the 23-24 segment, the $\psi$ torsion of D23 and $\phi$ torsion of V24 showed a similar PES curve. The slightly larger energy increase in the $\phi$ torsion of V24 than in the $\psi$ torsion of D23 arises from the difference in the $\Delta \rm N_{\rm HB}$.

The PESs for the $\psi$ torsion of G25 and the $\phi$ torsion of S26 torsional pathways showed a similar energy curve in the early part of the pathway. However, in the latter part, the PESs showed different energy curves; in the $\psi$ torsion of G25, the curve became relatively flat, while in the $\phi$ torsion of S26, the energy continued to increase. We observed that in the early part of the pathway, both torsions involved the breakage of the four mainchain HBs. In the latter pathway of the $\phi$ torsion of S26, unlike the $\psi$ torsion of G25, additional breakage of HB occurs in the 25-26 segment. This additional HB breakage likely contributes to the further increase in energy observed in the PES for the $\phi$ torsion of S26.

Next, we discuss the activation barrier that appears on the PES. 
The activation barrier for twisted-to-fibril conformational transition represents the propensity of the former to revert to the latter.
For each PES, an activation barrier for the twisted-to-fibril conformational transition was confirmed and summarized in Table 1. 
The height of these barriers ranged from approximately 0.4 to 0.6 eV, except for the $\psi$ torsion of G25, which exhibited a significantly lower barrier of only 0.05 eV.

To understand the origin of the activation barrier, we analyzed the structures at the activation barrier top and the local minimum on the PES.
At the top of the activation barrier, we found that some interatomic distances between an atom contained in the sidechain near the torsional axis and an atom in the lower monomer become characteristically close, with many increasing at the local minimum.
Table 1 lists these interatomic distances.
For the $\psi$ torsion of E22 and the $\phi$ torsion of D23, the hydrogen atoms of the E22 sidechain approach the hydrogen atoms in the lower monomer during the transition from the local minimum to the barrier top. 
For the $\psi$ torsion of D23 and the $\phi$ torsion of D24, the oxygen atoms of the carboxyl group in the D23 sidechain approach the mainchain carbonyl oxygen atom of the 22–23 segment in the lower monomer.

In comparison with these torsions, for the $\psi$ torsion of G25 and the $\phi$ torsion of S26, the sidechain near the torsional axis contains only a single hydrogen atom.
Therefore, the intermolecular steric repulsion encountered during the twisted-to-fibril conformational transition is greater in the $\psi$ torsion of E22, $\phi$ torsion of D23, $\psi$ torsion of D23, and $\phi$ torsion of V24 than in the $\psi$ torsion of G25 and the $\phi$ torsion of S26. 
This steric hindrance accounts for the barrier height observed for the $\psi$ torsion of E22, $\phi$ torsion of D23, $\psi$ torsion of D23, and $\phi$ torsion of V24.

We next explain why the $\phi$ torsion of S26 still exhibits an activation barrier despite the little intermolecular steric repulsion. At the local minimum on the PES for the $\phi$ torsion of S26, an intramolecular HB forms between the mainchain carbonyl oxygen atom of the 25–26 segment and a hydrogen atom of the S26 sidechain. The stabilized twisted conformation due to this HB likely creates the barrier height, compensating for the energy increase in the latter part of the PES for the $\phi$ torsion of S26.

\section{Discussion}

The PES for each torsional interconversion of the A$\beta$ exhibited an activation barrier for the twisted-to-fibril conformational transition.
Torsion occurring near the glycine, having a small sidechain (a single hydrogen atom), exhibited little steric hindrance and a low activation barrier. In contrast, torsion occurring near glutamic acid and aspartic acid, which possess relatively bulky sidechains, resulted in higher barriers due to greater steric hindrance. 
These results suggest that the size and orientation of the amino acid sidechain near the torsional axis strongly affect the barrier height and, thus, the tendency of the twisted conformation to revert to the fibril conformation.
We remark that other factors, such as the formation of intramolecular HB, as observed in the $\phi$ torsion of S26, can also influence this tendency. We exclude such a case in the following discussion.

The simulation in the present study does not consider the presence of the solvent around the A$\beta$.
Here, we discuss the possible effect of the solvent on the PES for the torsional motion of the A$\beta$.
The torsional transition from the fibril conformation increases the number of sites in the A$\beta$ where water molecules can form an HB. 
Thus, we suppose that, in the solution, the stability of the twisted conformation is improved, thereby lowering the endothermic energy for the fibril-to-twisted conformational transitions.
If the presence of solvent molecules does not greatly disrupt the local conformation around the torsional axis, the activation barrier originating from the steric hindrance is expected to be maintained.
We assume that the solvent effect is not strong enough to disrupt the local conformation around the torsional axis.
Under such conditions, the intermolecular steric repulsion caused by the sidechain near the torsional axis is encountered during twisted-to-fibril conformational transitions, allowing the twisted conformation to appear as local minima in the solution environment.
A similar discussion has been presented on another molecular system, where two metastable states interconverting via torsional motion retain an activation barrier in the solution environment, although their relative stability changes due to the hydration effect [39,40].

Suppose that the twisted conformations at the fibril ends are experimentally observed. 
In this case, the observed twisted conformation may be the one involving intermolecular steric hindrance at the residues having moderately bulky sidechains, rather than at the residues having small sidechains, such as glycine. The experimental validation of the present theoretical study is essential to further elucidate the structure and dynamics of the A$\beta$ at the fibril end.

\section*{Conclusions}
We theoretically investigated the structure and energetics of A$\beta$ at the fibril end. Several locally stable structures of twisted conformations, in which a part of the peptide chain is directed upward due to torsion around a specific single bond, were identified by DFT-based structural optimization.

All fibril-to-twisted conformational transitions were endothermic. In the fibril conformation, closely formed HBs between monomers contribute to its stabilization. The transition to the twisted conformation partly breaks HBs, and the degree of destabilization (endothermic energy) is correlated with the number of broken HBs. In the $\phi$ torsion of D23, the repulsion of the carbonyl oxygen atoms between the upper and lower monomers destabilizes the twisted conformation. The van der Waals interactions, such as $\pi$-stacking and other intermolecular and intramolecular contacts of the hydrophobic sidechains, likely stabilize the fibril conformation, and the loss of these interactions increases the endothermic energy required for the transition to the twisted conformation.

The PES connecting the fibril and twisted conformation was computed using the NEB method.
The activation barrier for twisted-to-fibril conformational transitions was identified, representing the propensity of the former to revert to the latter.
We compared the activation barriers for twisted-to-fibril conformational transitions.
Torsional motions involving the intermolecular steric hindrance caused by the sidechain near the torsional axis increase the activation barrier. The additional formation of HB also increases the activation barrier. Excluding such a case, the propensity of a twisted conformation to revert to the fibril conformation depends strongly on the local conformational environment around the torsional axis, such as the sidechain size and orientation. These findings provide a fundamental understanding of A$\beta$ conformation at the fibril end, which is crucial for predicting their structure and dynamics.

\section*{Conflicts of interest}
No conflicts of interest to declare.

\section*{Acknowledgments}
The authors thank Prof. M. Otani, Prof. N. Nakamura, and Prof. S. Hagiwara for fruitful discussions.
Y.O., M.K., and K.K. thank K. Komatsu and M. Fukui for the valuable comments on this research.
The calculations were implemented at the computer center of Kyushu University, ISSP of the University of Tokyo, and the Information Technology Center of the University of Tokyo.
This work was partially supported by the Japan Society for the Promotion of Science (JSPS) KAKENHI (Grant No. 24KJ1912).
Y.O. gratefully acknowledges the fellowship support from the JSPS.


\begin{figure}[h]
\centering
  \includegraphics[height=18cm]{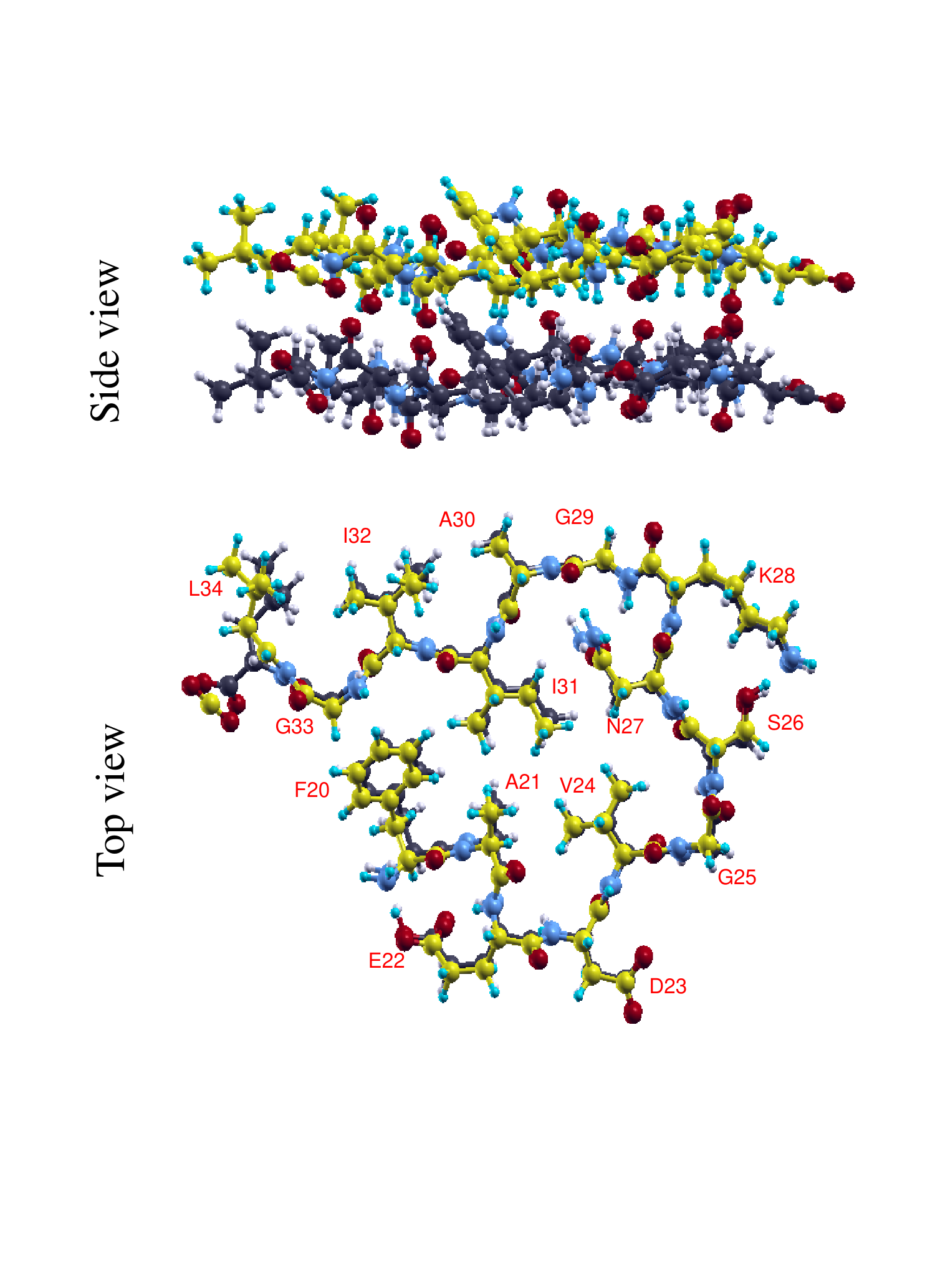}
  \caption{Side and top views of the optimized structure of the fibril conformation. Throughout this paper, the carbon and hydrogen atoms in the upper monomer are shown as yellow and cyan spheres, respectively, while the carbon and hydrogen atoms in the lower monomer are shown as black and white spheres, respectively. The nitrogen and oxygen atoms are shown as blue and red spheres, respectively. The amino acid label was added as red letters near the corresponding sidechains. Amino acid residues are denoted using the standard single-letter code with their position in the A$\beta$(1-42) sequence (e.g., F20 refers to phenylalanine at position 20) throughout this paper.}
  \label{fgr1}
\end{figure}

\begin{figure}[h]
 \centering
 \includegraphics[height=22cm]{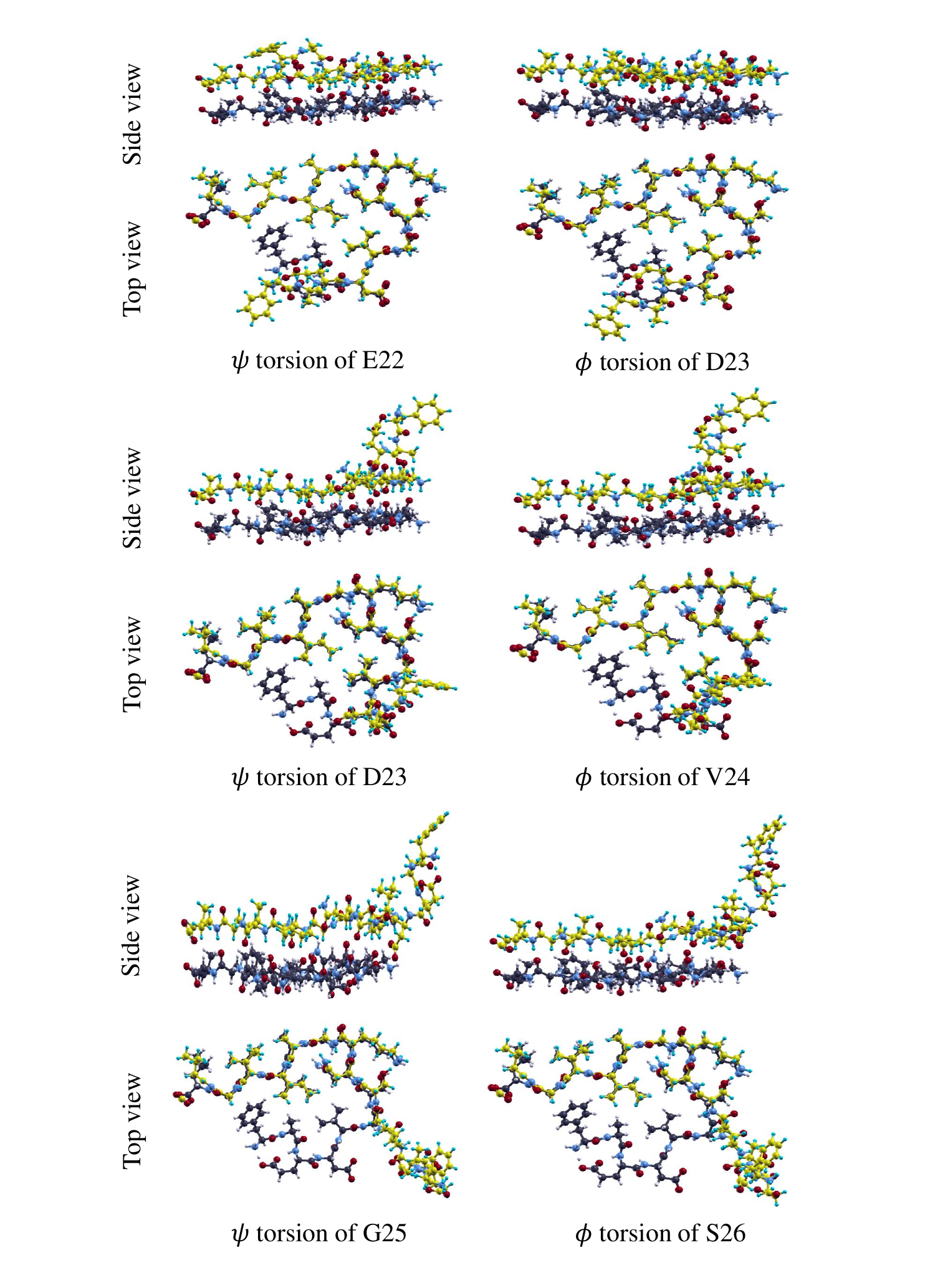}
 \caption{Side and top views of the optimized structure of the twisted conformation for each torsion.}
 \label{fgr2}
\end{figure}

\begin{figure}[h]
 \centering
 \includegraphics[height=8cm]{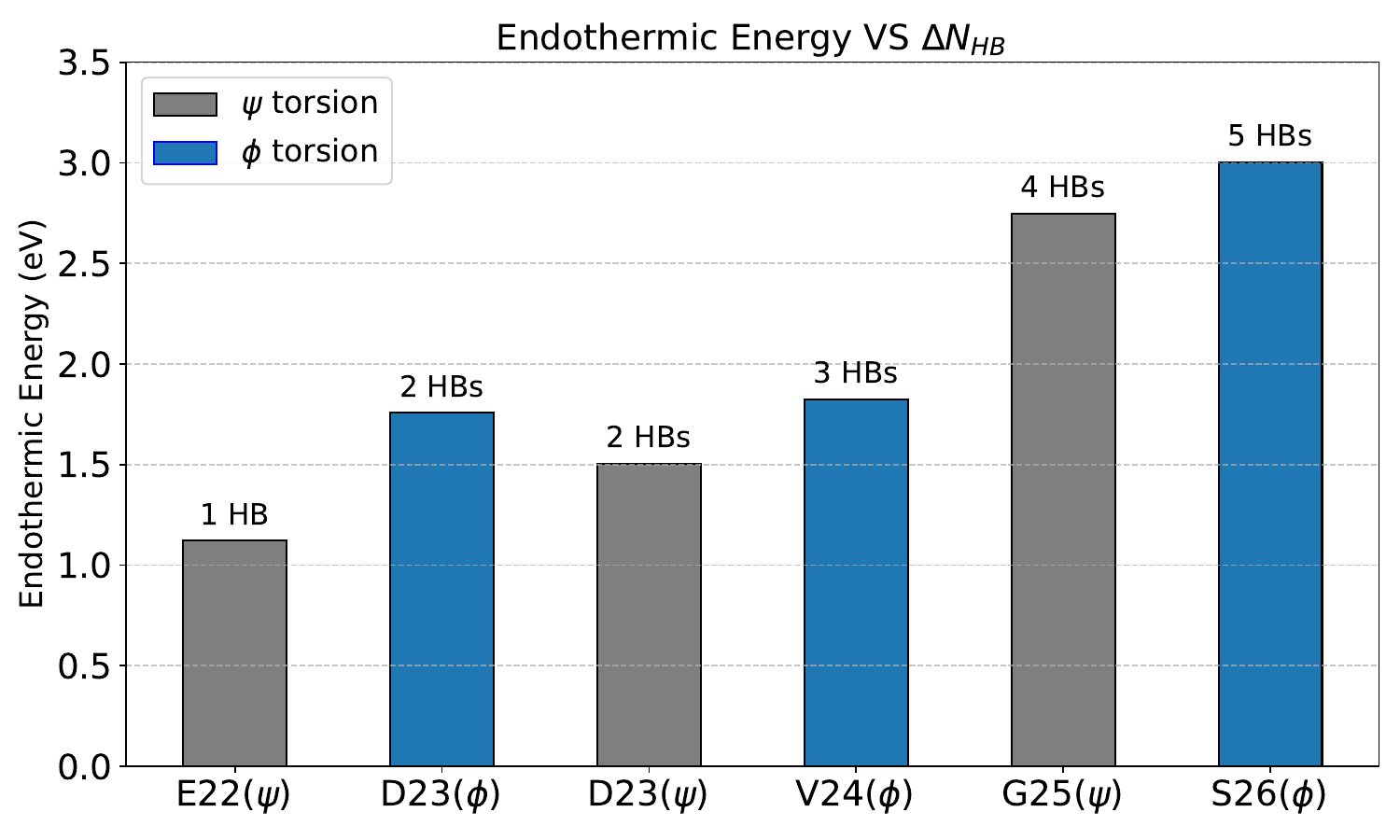}
 \caption{Energy of the twisted conformation relative to the fibril conformation for each torsion. The amino acid residue shown on the horizontal axis corresponds to the residue containing the torsional axis. The gray and blue bars represent the energies resulting from $\psi$ and $\phi$ torsions, respectively. On top of each bar graph, the number of mainchain hydrogen bonds broken due to each torsion ($\Delta \rm N_{\rm HB}$) is indicated.}
 \label{fgr3}
\end{figure}

\begin{figure}[h]
 \centering
 \includegraphics[height=22cm]{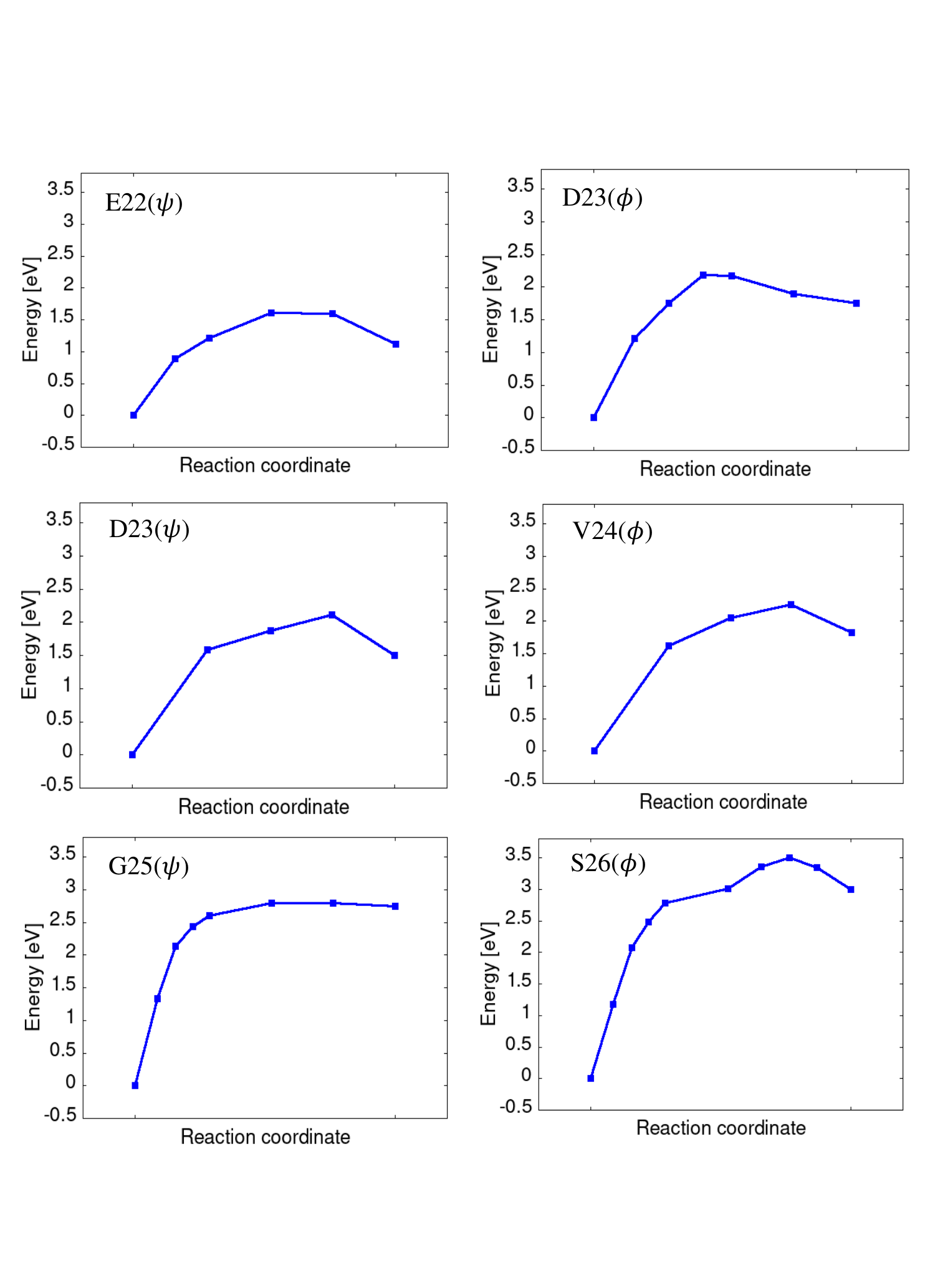}
 \caption{Calculated potential energy surface for each torsion. The energy of the fibril conformation was set as a reference (zero energy).}
 \label{fgr4}
\end{figure}

\begin{table*}[h]
\small
  \caption{The activation barriers from the twisted to the fibril conformation for each torsion are summarized. 
  We also present the distances ($d$) between atoms in the upper and lower monomers, which are considered to be related to intermolecular steric hindrance, at both the activation barrier top and the local minimum for each torsional pathway.
  For the $\psi$ torsion of E22 and the $\phi$ torsion of D23, the distances between the hydrogen atom contained in the E22 sidechain and the hydrogen atom in the lower monomer are shown.
  For the $\psi$ torsion of D23 and the $\phi$ torsion of D24, the distances between the oxygen atoms of the carboxyl group in the D23 sidechain and the mainchain carbonyl oxygen atom of the 22–23 segment are shown.
  For the $\psi$ torsion of G25 and the $\phi$ torsion of S26, the distance between one of the two hydrogen atoms bonded to the $\alpha$-carbon and a hydrogen atom in the lower monomer is shown (Note that the atom pair differs between the barrier top and the local minimum).
  }
  \label{tbl:example2}
  \begin{tabular*}{\textwidth}{@{\extracolsep{\fill}}lllllll}
    \hline
    Types of torsion & Activation barrier (eV) & Atom pair & $d$ at the barrier top (\AA) & $d$ at the local minimum (\AA) \\
    \hline
    E22($\psi$) & 0.49 & H-H & 1.7568 & 2.7152  \\
                       &  & H-H & 1.7432 & 1.9845  \\
    D23($\phi$) & 0.43 & H-H & 1.6629 & 5.9567   \\
                    &  & H-H & 1.9430 & 4.7694  \\
    D23($\psi$) & 0.61 & O-O & 2.6464 & 3.1902 \\
                    &  & O-O & 2.8930 & 3.0023  \\
    V24($\phi$) & 0.43 & O-O & 2.2776 & 2.9519   \\
                    &  & O-O & 2.9646 & 2.9632  \\
    G25 ($\psi$) & 0.05 & H-H & 2.8713 & 3.1348 \\
    S25 ($\phi$) & 0.50 & H-H & 2.3406 & 2.5905 \\
    \hline
  \end{tabular*}
\end{table*}

\end{document}